\documentclass[twocolumn,english,aps,prl,showpacs]{revtex4-1}
\usepackage{amssymb}
\usepackage{amsmath}
\usepackage[pdftex]{graphicx}

\begin{document}

\title{One-dimensional Quasicrystals from Incommensurate Charge Order}

\author{Felix Flicker}
  \email{flicker@physics.org}
  \affiliation{Department of Physics, University of California, Berkeley, CA 94720, USA}
\author{Jasper van Wezel} 
  \email{vanwezel@uva.nl}
  \affiliation{Institute for Theoretical Physics, University of Amsterdam, 1090 GL Amsterdam, The Netherlands}

%Charge-density-wave systems, Quasicrystals, Incommensurate crystals

\begin{abstract}
Artificial quasicrystals are nowadays routinely manufactured, yet only two naturally-occurring examples are known. We present a class of systems with the potential to be realised both artificially and in nature, in which the lowest energy state is a one-dimensional quasicrystal. These systems are  based on incommensurately charge-ordered materials, in which the quasicrystalline phase competes with the formation of a regular array of discommensurations as a way of interpolating between incommensurate charge order at high temperatures and commensurate order at low temperatures. The nonlocal correlations characteristic of the quasicrystalline state emerge from a free energy contribution localized in reciprocal space. We present a theoretical phase diagram showing that the required material properties for the appearance of such a ground state allow for one-dimensional quasicrystals to form in real materials. The result is a potentially wide class of one-dimensional quasicrystals. 
\end{abstract}

\maketitle

While many quasicrystals have been artificially created since their first synthesis in 1984, only two naturally-occurring quasicrystals have been identified to date -- both found in the same Siberian meteorite~\cite{Shechtman84,Steinhardt09,Steinhardt15}. In the quasiperiodic, never-repeating, pattern of two or more unit cells defining a quasicrystal (QC)~\cite{SteinhardtSocolar,Senechal}, locally changing a single cell may require adjustment of an infinite number of other cells in order to maintain the quasiperiodicity~\cite{Achim14}. This global property makes it non-trivial to grow a QC cell-by-cell. Proposed mechanisms in two or higher dimensions include recognising vertex matching rules which imply `forced tiles', random tiling models, and relaxation processes from non-quasiperiodic systems~\cite{Janot}. 

\begin{figure}
\centerline{{\includegraphics[width=0.98 \columnwidth]{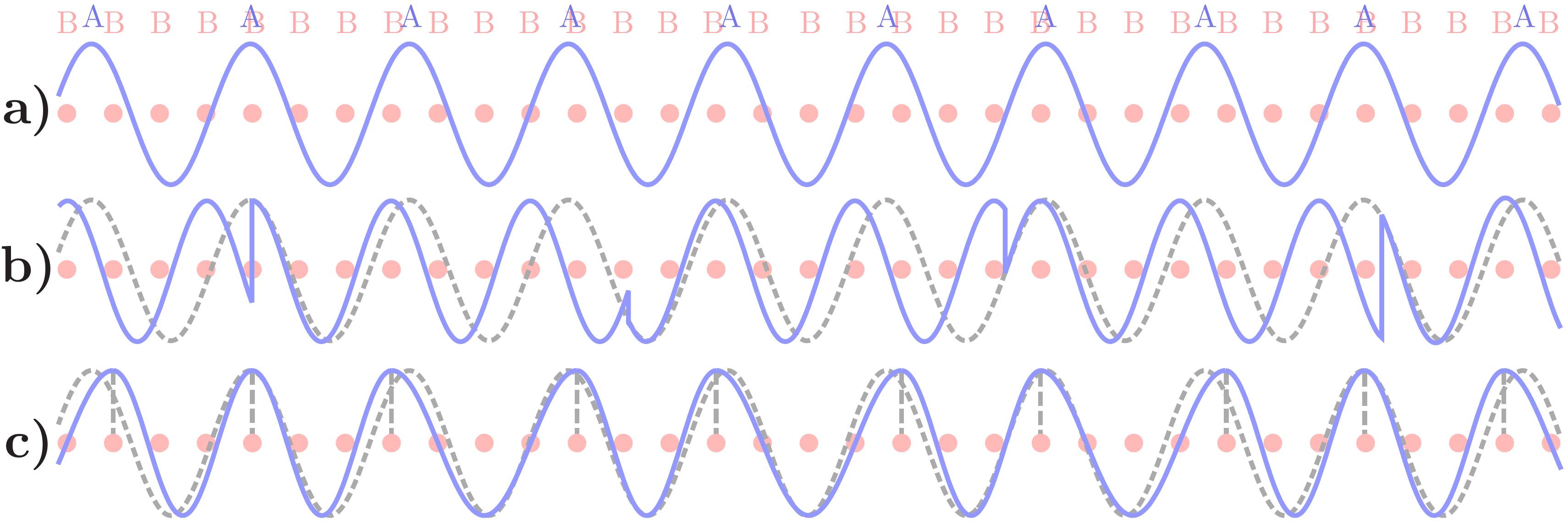}}}
\caption{(color online) Possible charge orderings when the electron density (blue) develops an instability with wavevector $Q$ incommensurate with the ion lattice (pink). Here we used a wavelength of $2\pi/Q=\sqrt{11.7}\approx3.42$ lattice spacings. {\bf (a)} Perfectly incommensurate charge order. Labeling the blue peaks A and the pink ions B, the sequence of A's and B's is a quasiperiodic tiling of the units ABBB and ABBBB~\cite{us}. {\bf (b)} Discommensuration state, which is locally commensurate with period 3, but maintains the average wavevector $Q$~\cite{McMillan76}. {\bf (c)} A quasicrystal composed of a quasiperiodic arrangement of period $3$ and period $4$ unit cells, maintaining the average wavevector $Q$. This state is reached by shifting each peak A onto the nearest ion B. Dotted lines show the incommensurate state for comparison. In all cases the ion positions will adjust to the altered electronic charge distribution (not shown). An alternative comparison between these states is provided in the Supplementary Material.
}
\label{mCDW}
\end{figure}

In (quasi) one-dimensional (1D) systems, QCs consist of a quasiperiodic sequence of long and short unit cells. The global nature of 1D QCs is apparent from their correspondence to projections of regular 2D crystals~\cite{SteinhardtSocolar,ourEPL}. One way to experimentally access properties of 1D QCs is to consider instead a setup containing two periodic subsystems that are incommensurate with respect to one another. The sequence of subsystem species encountered as one traverses the system is aperiodic, and can be mapped onto a similar aperiodic or quasicrystalline sequence of unit cells (see Fig.~\ref{mCDW}(a)). Such incommensurate systems have recently been employed in a number of experimental studies of QCs: optical waveguide arrays simulating 1D crystals with incommensurate periodic on-site potentials; cold atom condensates in optical lattices;  Moir\'{e} superlattices in graphene derivatives; and in theoretical descriptions of incommensurate charge density waves~\cite{Kraus1,Kraus2,Bloch,Moire,ourEPL}.  While these are of much interest in their own right, they only constitute quasicrystals by way of mathematical analogy. One key difference with true quasicrystals for example, is that incommensurate systems have no lower bound on the spatial separation between sites of the two subsystems; it is only the sequence which is quasiperiodic (see also the Supplementary Material).

In this Letter we describe a mechanism for generating 1D quasicrystals based on coupling two incommensurate subsystems and allowing them to adjust to minimize their overall free energy. We take as a specific example the case of incommensurate charge density waves, showing that, under specific conditions, it is energetically favourable for the atomic lattice to adopt a quasiperiodic structure. The resulting state is thus a true quasicrystal, transcending the mere mathematical equivalence between 1D QCs and incommensurate charge density modulations on top of a periodic atomic background. We show that the required material properties are physically realistic. Finally, we outline the criteria needed for this mechanism to apply more generally.

Low-dimensional materials are prone to develop charge order. In the ideal 1D case, a Peierls instability always yields a spontaneous periodic modulation of the electron density, accompanied by periodic displacements in the atomic lattice with the same wavevector~\cite{Peierls}.
In real quasi-1D (or higher-dimensional) materials, charge order typically occurs only if the electron-phonon coupling is sufficiently strong~\cite{ChanHeine,us}. In that case the order arises when the electronic susceptibility overcomes the cost of populating phonon modes, so that:
\begin{align}
\int d{\bf k} ~ |g({\bf k},{\bf k+q})|^2 \frac{f(\epsilon_{\bf k}) - f(\epsilon_{\bf k+q})}{\epsilon_{\bf k+q} - \epsilon_{\bf k}} \ge \hbar \omega_{\bf q}
\label{susc}
\end{align}
where $g({\bf k},{\bf k+q})$ is the electron-phonon coupling, which is generically momentum- and orbital-dependent. $f(\epsilon_{\bf k})$ is the Fermi-Dirac distribution for an electron with momentum ${\bf k}$ and energy $\epsilon_{\bf k}$, and $\hbar \omega_{\bf q}$ is the energy of the phonon mode which softens in the charge ordering transition. The full susceptibility $\chi({\bf q})$, defined by the left-hand side of Eq.~\eqref{susc}, has its maximum at $\tilde{\bf q}$, which determines the wavevector $\mathbf{Q}(T_{\text{CDW}}) = \tilde{\bf q}$ at which the charge order and atomic displacements first form. In the presence of strong nesting, guaranteed in 1D, $\tilde{\bf q}$ equals the nesting vector. In higher dimensions it is determined by the momentum dependences of both the electron-phonon coupling and the electronic dispersion~\cite{us}.

The evolution of the charge density's wavevector ${\bf Q}(T)$, as temperature is lowered beyond the onset temperature $T_{\text{CDW}}$, can be modeled by a Ginzburg-Landau expansion of the free energy~\cite{McMillan75,Yejun}:
\begin{align}
F = &-\int d{\bf q} ~ \chi({\bf q}) \rho^2({\bf q}) \notag \\
 & +  \int d{\bf x} ~ b \rho^4({\bf x}) - \sum_{n,{\bf K}} c_n \cos({\bf K}\cdot{\bf x}) \rho^n({\bf x}).
\label{F}
\end{align}
The modulation on top of the average electron density is written as $\rho(\mathbf{x}) = \psi \cos( \phi(\mathbf{x}) )$, where for a density wave without defects $\phi(\mathbf{x}) = \mathbf{Q} \cdot \mathbf{x}$. For notational convenience the quadratic part of the free energy has been written in reciprocal space, where it is proportional to the full electronic susceptibility $\chi$. This term favors the formation of charge order at wavevector $\tilde{\mathbf{q}}$. The other terms are more conveniently written in real space. Those proportional to $c_n$, with $n>2$, represent the local coupling between the atomic lattice and the charge modulations. These favor charge modulations locally commensurate with the lattice, by giving a non-zero contribution to $F$ whenever $n \phi(\mathbf{x}) = \mathbf{K}\cdot \mathbf{x}$, with $\mathbf{K}$ a reciprocal lattice vector of the atomic lattice~\cite{McMillan75,McMillan76}.

At the transition temperature $T_{\textrm{CDW}}$, the order parameter amplitude $\psi$ is small, so that the lower-order terms in Eq.~\eqref{F} dominate, and the order forms at wavevector $\mathbf{Q}(T_{\textrm{CDW}}) = \tilde{\mathbf{q}}$. As temperature is lowered, $\psi$ increases, and the terms proportional to $c_n$ begin to compete with the second-order term. Since the coupling to the lattice has an effect only for the parts of the density wave that are locally commensurate, the charge order with $\phi(\mathbf{x}) = \mathbf{Q}\cdot \mathbf{x}$ will not gain energy from the final terms for any incommensurate value of $\mathbf{Q}$. Instead, the charge-ordered state can lower its energy by adopting a locally commensurate structure $\phi(\mathbf{x}) = \mathbf{K}\cdot \mathbf{x}-\delta(\mathbf{x})$, where $\delta(\mathbf{x})$ introduces a regular array of broadened phase slips (discommensurations)~\footnote{We employ the standard assumption of the array being periodic. Although strong lattice coupling may in principle alter the distribution, this will not affect the qualitative difference between the discommensuration state and any of the other states discussed here. Only the character of the transition between them may be altered.}, which render the average propagation vector incommensurate even if the local structure is predominantly commensurate~\cite{McMillan76}. The evolution as temperature decreases further is characterized by a sharpening of the initially-broad discommensurations~\cite{McMillan76,Yejun}. 

In many materials, $\mathbf{Q}(T)$ starts out incommensurate at the onset of the charge-ordered state, evolves towards a close-by commensurate value, and locks in at a second transition temperature $T_{\textrm{IC}}$ below which the charge order remains commensurate. Well-known examples can be seen in $2H$-TaSe$_2$ and TTF-TNQ~\cite{Kagoshima76,Moncton77}. In other materials, the evolution of $\mathbf{Q}(T)$ starts out in a similar manner, but remains incommensurate to zero temperature. Examples include K$_{0.3}$MoO$_3$, TbTe$_3$, and NbSe$_2$~\cite{Moncton77,Moudden90,Ru08}. A proliferation of discommensurations can describe either type of behavior, depending on the value of the ratio $c_n / \chi(\tilde{\bf q})$ in the free energy of Eq.~\eqref{F}, which determines the strength of the coupling to the lattice relative to the maximum strength of the full susceptibility. The presence of discommensurations in these materials has been experimentally verified, for example in scanning tunneling microscopy experiments on NbSe$_2$~\cite{Soumyanarayanan}.

The discommensuration state takes advantage of all the terms in the free energy of Eq.~\eqref{F}. Nevertheless, one can imagine conditions under which it may not be the optimal configuration of $\phi(\mathbf{x})$. Consider incommensurate charge order which forms at $T_{\text{CDW}}$ with an incommensurate wavelength $\lambda = 2 \pi / \tilde{q}$ lying close to halfway between three and four lattice spacings (Fig.~\ref{mCDW}(a)). In that case, it is not \emph{a priori} obvious whether the coupling to the lattice will prefer the charge order to lock in at period three or four. Choosing one, and introducing discommensurations such that the local electronic density modulations are commensurate while the average propagation vector remains incommensurate, necessarily results in a high density of rapid phase changes (shown in Fig.~\ref{mCDW}(b)). These sharp discommensurations involve high momentum components of $\rho(\mathbf{q})$ (higher harmonics of $\tilde{\mathbf{q}}$), and are therefore energetically costly with respect to the quadratic term in $F$. An alternative way of distorting the incommensurate charge order and gaining local lock-in energy, while retaining the overall incommensurability, is depicted in Fig.~\ref{mCDW}(c). This configuration is obtained by moving every maximum in the charge density wave directly onto the atomic position closest to it. The result is a charge modulation that is locally commensurate everywhere, but with two local periodicities. The locally-commensurate patches are smoothly connected (although higher derivatives are not smooth), and therefore do not involve the high-momentum components characteristic of the discommensurations. For a more detailed description of the difference between this state and the discommensuration state, see the Supplementary Material.

It is clear from the incommensurability that the pattern of period three and four modulations in the multiply-commensurate structure never repeats itself. In fact it can be shown, using a cut-and-project construction, that the arrangement is quasiperiodic~\cite{us}. In reality the ions will be displaced towards the charge maxima simultaneously with the displacement of the charge maxima towards the ions. The structure shown in Fig.~\ref{mCDW}(c) will therefore not just contain incommensurate charge modulations or a quasiperiodic electronic structure, but will constitute a true 1D quasicrystal. At the onset temperature of the multiply-commensurate order, there will be a crystal-to-quasicrystal phase transition breaking all translational symmetries of the original lattice.
\begin{figure}
\centerline{{\includegraphics[width=0.98 \columnwidth]{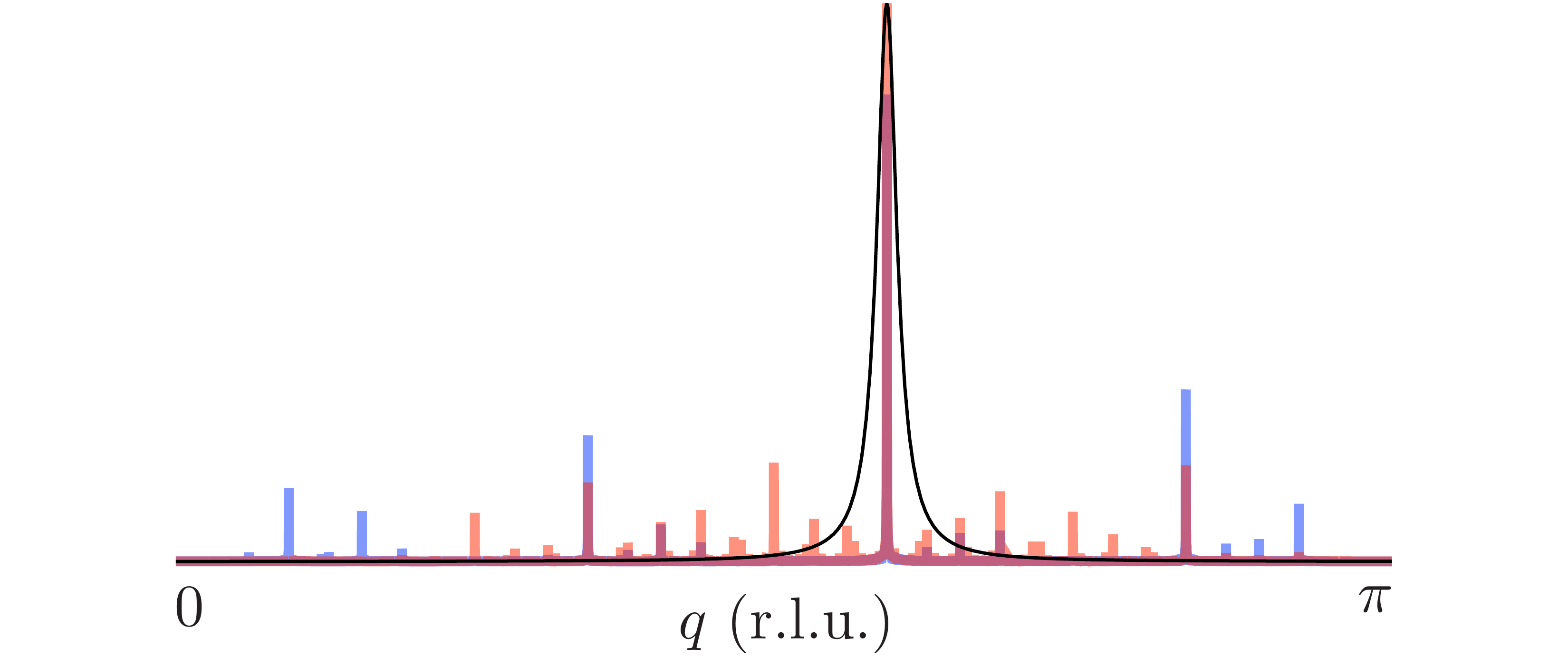}}}
\caption{(color online) Calculated X-ray diffraction pattern for the discommensuration state (blue) and quasicrystal (pink) shown in Fig.~\ref{mCDW} (purple indicates overlapping peaks). The black line depicts the approximation to the full susceptibility $\chi$ used to calculate the phase diagram of Fig.~\ref{phases}. Both types of order have their main peak at $\tilde{\mathbf{q}}$, but the quasicrystal has a dense covering of secondary peaks close to the main diffraction spot, which benefit from the nonzero $\chi$ in that region. The units are arbitrary, but both diffraction patterns have the same scale.
}
\label{Fourier}
\end{figure}

The X-ray diffraction pattern of a quasicrystal is characterized by a dense array of sharp peaks~\cite{Senechal,Janot}. As shown in Fig.~\ref{Fourier}, the Fourier transform of the multiply-commensurate charge order indeed contains such an arrangement, centered around the average incommensurate wavevector $\tilde{\mathbf{q}}$. Real quasi-1D materials have susceptibilities $\chi(\mathbf{q})$ which decay over a finite range of momenta. The clustered peaks around $\tilde{\mathbf{q}}$ can therefore also gain energy from the quadratic term in $F$. This situation should be contrasted with that of the Fourier transform of the discommensuration state, also shown in Fig.~\ref{Fourier}. In that case, although the central peak gains energy, the higher harmonics typically lie far away in $k$-space, and cannot utilize the non-zero width of the susceptibility. The difference in the extent to which these structures profit from the quadratic term in $F$ can render the quasicrystal energetically favourable to the discommensuration state.

\begin{figure}
\centerline{{\includegraphics[width=0.98 \columnwidth]{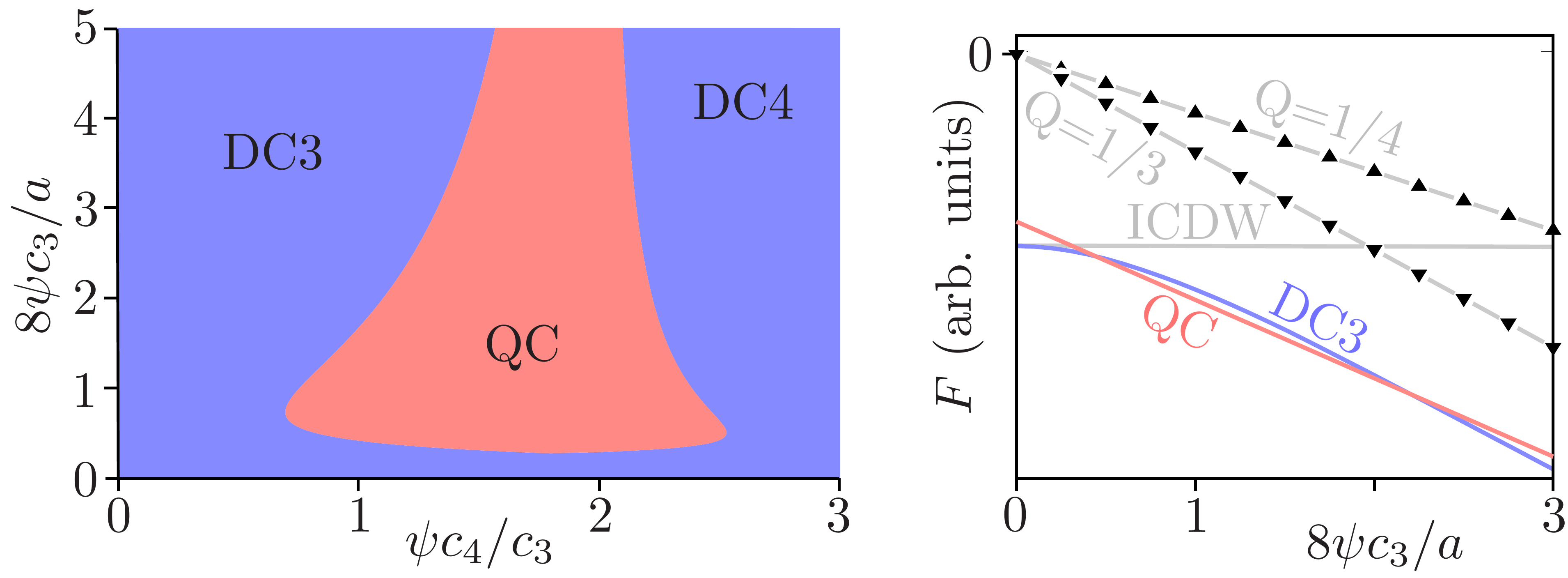}}}
\caption{(color online) {\bf (a)} the predicted phase diagram emerging from Eq.~\eqref{F}, for $Q=2\pi / \sqrt{11.7}$ and $b=1$. {\bf (b)} the free energies of the different ordered states as a function of $\psi c_3/a$ for fixed $\psi c_4/c_3 = 1.2/8$. The abbreviations used are: QC--quasicrystal, DC$n$--discommensuration state with commensurate areas of period $n$, ICDW--incommensurate charge density wave, and $Q=1/n$--commensurate charge order with period $n$.
}
\label{phases}
\end{figure}

To quantify the relative stabilities of the quasicrystal and discommensuration states, we consider again the free energy of Eq.~\eqref{F}. We model the full susceptibility as $\chi(\mathbf{q}) = a / (\sigma^2 (|\mathbf{q}|-\tilde{\mathbf{q}})^2+1)$, where the width $1/{\sigma}$ of the peak is taken to be small but non-zero. The ratio $b/a$ of the quartic and quadratic component of $F$ at $\mathbf{q}=\tilde{\mathbf{q}}$ determines the overall magnitude of the order parameter $\psi$, but does not affect the form of $\phi({\mathbf{x}})$. Whether the discommensuration or quasicrystalline state is favorable is thus determined by the ratios $(\psi^4 c_4)/(\psi^3 c_3)$ and $(\psi^3 c_3)/(\psi^2 a)$. In Fig.~\ref{phases}(a) we present the phase diagram as a function of these two ratios. The energy of the quasicrystalline state is compared to that of an array of discommensurations with optimized density and widths. A comparison of the energies of individual states, as a function of $\psi c_3/a$ for fixed $\psi c_4/c_3$, is shown in Fig.~\ref{phases}(b). 

The energy of the incommensurate state does not depend on the coupling to the lattice, and is constant in $\psi c_3/a$. The commensurate state on the other hand does not gain any energy from the susceptibility term, but takes full advantage of the lock-in term, and decreases linearly in energy as $\psi c_3/a$ increases. At high coupling to the lattice the commensurate state thus becomes favorable to the incommensurate. The discommensuration state interpolates between these two extremes, having its main Fourier component at the incommensurate value, but being locally commensurate. It takes advantage of all terms in the free energy, and outperforms both the commensurate and incommensurate states.

The quasicrystal also has its dominant Fourier component at the incommensurate value. At intermediate values of $\psi c_3/a$ it additionally takes advantage of the spread in $\chi(\mathbf{q})$ and only pays a modest price for not being optimally commensurate everywhere. In this regime, the quasicrystal is favored over the discommensuration state, giving rise to the extended area of stability shown in Fig.~\ref{phases}(a). At either extreme of $\psi c_3/a$, however, the discommensurations win out: the spectral weight of the dominant QC Fourier peak is constant, so does not approach the weight of the incommensurate state in the limit of zero lattice coupling. The discommensuration state is therefore favorable to the quasicrystal at very low values of $\psi c_3/a$. For strong coupling to the lattice, the locally-commensurate nature of the quasicrystal allows it to gain lock-in energy, with the overall energy gain a weighted average of the $c_3$ and $c_4$ terms. However, if these two coefficients are not equal, the quasicrystal cannot gain as much energy as the discommensuration state, which is always locally commensurate with the most favorable commensurability. The discommensuration state therefore also wins out over the quasicrystal at very high values of the coupling to the lattice.

Based on these considerations, the prerequisites for this type of 1D quasicrystal to emerge in any real (natural or artificial) material are: (i) a quasi-1D charge-ordered material with an incommensurate wavevector lying close to halfway between two commensurate values; (ii) a sharp, but not infinitely sharp, susceptibility; (iii) a strong, momentum-independent electron-phonon coupling. The latter is necessary to ensure that the coupling of the charge order to the lattice does not strongly favor any particular commensurate value of the propagation vector. 

Quasi-1D materials contain strongly-bonded chains of atoms with relatively weak inter-chain coupling. Their Fermi surfaces typically contain nearly-flat pieces, giving a strongly-peaked susceptibility. The charge ordering wavevector is therefore close to $2 k_{\text{F}}$, and is generically incommensurate. The question then arises as to why we are not overwhelmed with naturally-occurring quasicrystals in quasi-1D materials. First, the charge modulation $\rho$ is small compared to the average charge density, implying that a dense set of peaks neighboring $\tilde{\mathbf{q}}$ in the diffraction pattern may not be straightforward to distinguish experimentally. Second, in real materials it may be expected that the lock-in energy is different at different commensurate values~\cite{Yejun}, putting most materials close to the left side of Fig.~\ref{phases}(a). Note, however, that the order parameter amplitude $\psi$ increases with decreasing temperature, and thus enhances the value of $\psi c_4/ c_3$. Conversely, at the onset of ordering, $\psi$ is vanishingly small, and the free energy is dominated by the second-order term. This implies that the charge-ordered phase starts off at the origin of the phase diagram, and moves upwards and to the right only as temperature decreases below $T_{\textrm{CDW}}$. Third, except in cases of extremely strong electronic nesting, the momentum- and orbital-dependence of the electron-phonon coupling is generically essential for charge order to develop in real materials~\cite{us}. These points aside, the results of Fig.~\ref{phases} suggest that a targeted search may reveal previously-unnoticed quasicrystalline phases at low temperatures in quasi-1D materials with incommensurate charge order. Experimentally, the quasicrystalline charge order could be identified for example using microscopy to directly identify arrangements of neighbouring period-four and period-three cells, or using diffraction probes to find a fractal distribution of sharp peaks.

To conclude, we have shown that for an extended range of material parameters quasi-1D systems minimize their free energy by developing quasiperiodic tilings of two inequivalent unit cells. This provides a wide class of 1D materials with the potential to become quasicrystalline. Candidate materials, both natural and artificial, could be revealed by a directed search for materials exhibiting incommensurate charge order, with a wavelength close to halfway between two commensurate values, whose electronic structures are well-nested, and whose electron-phonon couplings are strong but momentum independent. The quasicrystalline state also extends the set of possible low-energy configurations of incommensurately charge-ordered systems, adding to the known incommensurate and commensurate states, and their interpolation via discommensurations.

Existing proposals for quasicrystal growth mechanisms have centered on single-cell additions at defect sites, or local adjustments based on propagating defects~\cite{Janot}. The quasicrystalline state discussed here, on the other hand, arises from a global adjustment of two periodic but incommensurate parent structures. The required co-ordination for this global effect in real space derives from the fact that it is a localized peak of the susceptibility in reciprocal space which determines the dominant wavevector of the quasicrystalline state.

Since the free energy contains two incommensurate periods, it is perhaps unsurprising that a ground state exists which respects the lack of translational symmetry. Closely related states have been found in both the quantum Hall effect in the Tao-Thouless limit~\cite{Bergholtz08}, and the formation of Wigner lattices in the Hubbard model~\cite{Hubbard}. In the latter case, the Wigner lattice is invoked to explain the optical spectra of nearly-quasicrystalline states in the 1D charge-ordered material TTF-TCNQ~\cite{Torrance75,Tanaka76}. Both systems, however, concern high-denominator rational commensurate periods rather than the incommensurate situation considered here, and neither depends on the particular free energy arguments arising from Eq.~\eqref{F}.

The mechanism for forming quasicrystals considered in this paper has the potential to arise whenever two incommensurate periodicities are simultaneously present in the same system. If an appropriate coupling between the subsystems can be introduced in the presence of a global constraint, a free energy similar to that of Eq.~\eqref{F} may be expected to apply, and a quasicrystalline state will generically emerge over an extensive portion of phase space.

\begin{acknowledgments}
{\bf Acknowledgements}---The authors wish to thank Shane Farnsworth for helpful discussions. JvW acknowledges support from a VIDI grant financed by the Netherlands Organisation for Scientific Research (NWO).
\end{acknowledgments}

\subsection{Supplementary Material}
\noindent
To bring to the fore the differences between the various charge-ordered phases 
discussed in the main text, it is convenient to express the modulation on top 
of the average charge modulation in each state as:
\begin{align}
\label{defphase}
\rho(\bf{x}) = \psi \cos\left({\bf K}\cdot{\bf x} - \delta({\bf x})\right).
\end{align}
Here, ${\bf K}$ is a wave vector commensurate with the lattice, which we will 
set to ${\bf K} = 2\pi/(3a)$ from here on. The commensurate charge density wave 
(CCDW), incommensurate charge density wave (ICDW), discommensuration state 
(DC), and quasicrystalline state (QC) can then be distinguished by considering 
the distinct behaviour of the phase $\delta({\bf x})$ in each. This is shown in 
Fig.~\ref{SMfig} below.

For the CCDW, $\delta({\bf x})$ is zero. For the ideal ICDW, $\delta({\bf x})$ 
is the straight line $\delta({\bf x})=({\bf Q}-{\bf K}) {\bf x}$, with ${\bf 
Q}$ the incommensurate wave vector. The discommensuration phase is given by 
sections of zero slope connected by vertical steps at regular intervals. The 
height of each step is such that the density wave jumps forward by precisely 
one lattice spacing. For ${\bf K}=2\pi/(3a)$, the step height is thus $2\pi/3$. 
The width of the horizontal sections is such that the average slope of the 
entire stepped structure is equal to ${\bf Q}$. The quasicrystal also has 
average slope ${\bf Q}$, but consists of horizontal sections separated by 
sections with slope $\delta({\bf x})=({\bf K'}-{\bf K}) {\bf x}$, where ${\bf 
K'}$ is a second commensurate wave vector: in this case $2\pi/(4a)$. The 
sections of zero and non-zero slope in this phase form a quasiperiodic sequence.

The DC phase often arises in real materials as the most 
energetically-favourable way of combining electronic density modulations with a 
lattice coupling~\cite{McMillan75}. In practice, the discommensurations in 
these materials will not be perfectly sharp. Instead, they broaden slightly and 
the connection in $\delta(\bf{x})$ with the neighbouring horizontal regions is 
made smooth. We follow McMillan in accounting for this using a Fourier expansion 
with coefficients, controlling the discommensuration width, selected so as to minimize the free energy of the DC state.
The QC phase, on the other hand, is created by starting from an ideal 
ICDW and then shifting the peaks of the electronic density modulation directly 
onto the nearest atomic position. This creates a quasiperiodic sequence 
consisting of sections with one of two well-defined commensurate wave vectors. 
\begin{figure}
\centerline{{\includegraphics[width=0.78 \columnwidth]{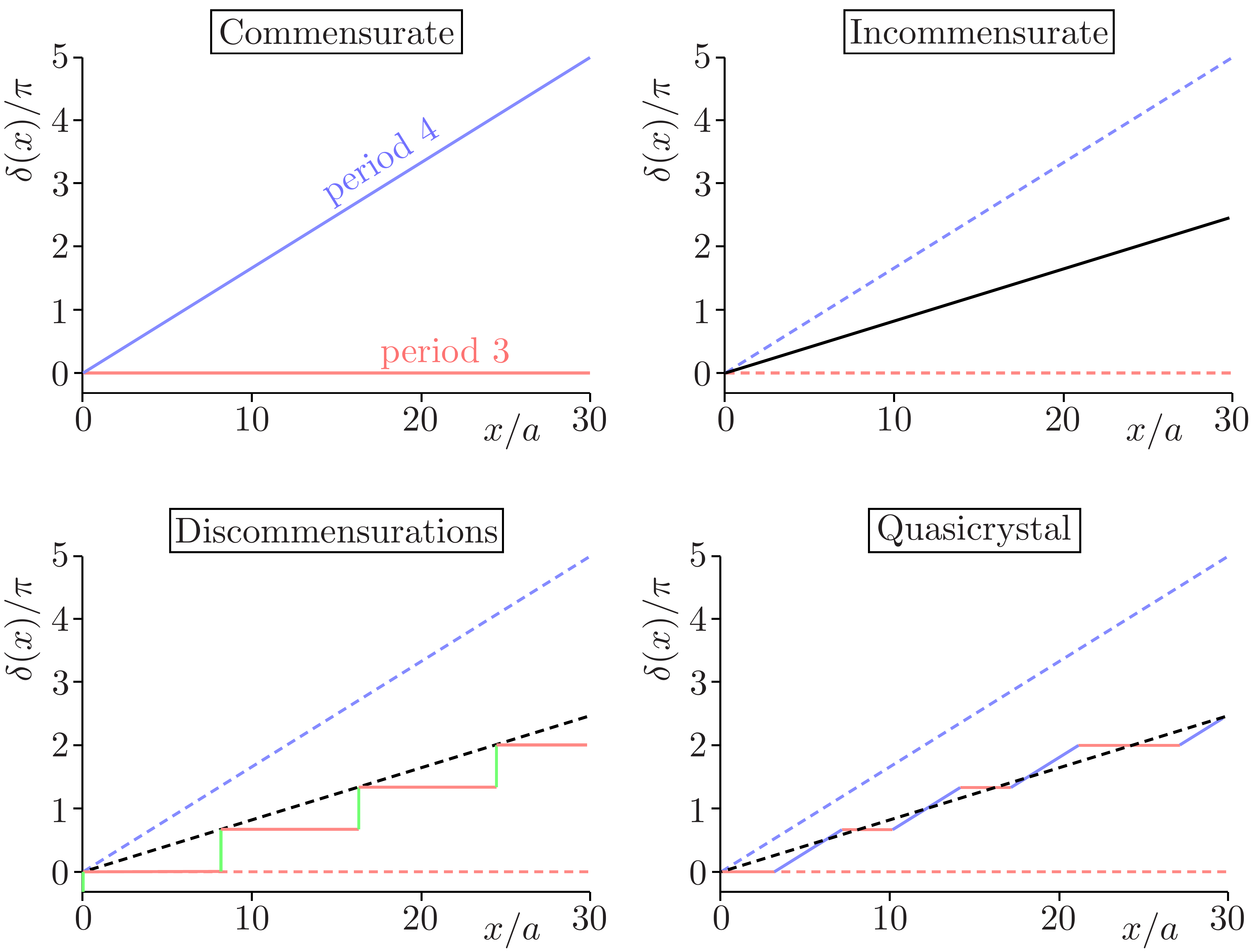}}}[h]
\caption{(color online) Possible charge orderings discussed in the main text. 
For all cases the phase $\delta({\bf x})$ defined in Eq.~\eqref{defphase} is 
shown divided by $\pi$, as a function of position in units of the lattice 
spacing $a$. With these conventions, the commensurate phases are represented by 
straight lines with rational values for their slopes. The incommensurate phase 
is a straight line with an irrational slope. The discommensuration phase 
consists of horizontal sections connected by a regular, periodic array of 
discommensurations, which are vertical lines of height $2/3$. The spacing of 
the discommensurations is such that the phase returns to the dashed 
incommensurate line at every discommensuration. Finally, the quasicrystal 
consists of a quasiperiodic sequence of horizontal (period $3$) sections of 
width $3$, and sections at the slope corresponding to commensurate order with 
period $4$, with width $4$. The aperiodic sequence can be constructed by always 
choosing the section whose end-point stays closest to the dashed incommensurate 
line.
}
\label{SMfig}
\end{figure}

The QC phase and ICDW both have quasiperiodic elements. For the ICDW, the 
sequence of atomic positions and charge maxima encountered as the material is 
traversed is quasiperiodic. This type of `quasiperiodicity', however, is 
generic to any superposition of two incommensurate structures, including even things like two parallel picket fences seen from a distance. In 
contrast, the QC phase is quasiperiodic in each of its components individually. 
Moreover, it also consists of precisely two unit cells. Each piece of period 
four within the QC state is locally indistinguishable from any other piece of 
period four. In the ICDW, the atomic neighbourhood of each peak in the 
electronic structure is perfectly unique. In that sense, it contains infinitely 
many unit cells, rather than two. We therefore classify the QC state as a true 
one-dimensional quasicrystal, in contrast to the ICDW, which is only a generic 
quasiperiodic sequence of charge maxima and atoms.

\end{document}